\documentclass[twocolumn]{aastex6}

\turnoffedit

\begin{document}

\title{\edit1{Observations of the Coronal Mass Ejection with a Complex Acceleration Profile}}

\author{A.A.~Reva, A.S.~Kirichenko, A.S.~Ulyanov, and S.V.~Kuzin}
       
\affil{Lebedev Physical Institute, Russian Academy of Sciences, Moscow, Russia;
                     email: \url{reva.antoine@gmail.com} }
             
\begin{abstract}

\edit1{We study the coronal mass ejection (CME) with a complex acceleration profile. The event occurred on April 23, 2009. It had} \edit2{an} \edit1{impulsive acceleration phase,} \edit2{an} \edit1{impulsive deceleration phase, and a second impulsive acceleration phase. During its evolution, the CME showed signatures of different acceleration mechanisms: kink instability, prominence drainage, flare reconnection, and a CME-CME collision.} The special feature of the observations is the usage of the TESIS EUV telescope. The instrument could image the solar corona in the Fe~171~\AA\ line up to a distance of 2~$R_\odot$ from the center of the Sun. This allows us to trace the CME up to the LASCO/C2 field of view without losing the CME from sight. \edit1{The onset of the CME was caused by kink instability. The mass drainage occurred after the kink instability. The mass drainage played only an auxiliary role: it decreased the CME mass, which helped to accelerate the CME. The first impulsive acceleration phase was caused by the flare reconnection. We observed the two ribbon flare and an increase of the soft X-ray flux during the first impulsive acceleration phase. The impulsive deceleration and the second impulsive acceleration phases were caused by the CME-CME collision. The studied event shows that CMEs are complex phenomena that cannot be explained with only one acceleration mechanism. We should seek a combination of different mechanisms that accelerate CMEs at different stages of their evolution.}

\end{abstract}

\section{Introduction}

\edit1{Coronal mass ejections (CMEs) are giant} \edit2{eruptions} \edit1{of the coronal plasma into interplanetary space. CMEs occur due to large releases of the magnetic energy on the Sun.} \edit2{Investigations} \edit1{of the mechanisms of the CME acceleration are important for solar physics and questions of the solar-terrestrial connections.}

\edit1{A lot of different models describe CME acceleration \citep{Chen2011}. The onset of the CME could be triggered by MHD instabilities \citep{Torok2005, Kliem2006}, mass drainage \citep{Fan2003, Reeves2005}, or breakout reconnection \citep{Antiochos1999}. The impulsive acceleration of the CME is thought to be} \edit3{connected with} \edit1{the flare reconnection \citep[CSHKP model, ][]{Carmichael1964, Sturrock1966, Hirayama1974, Kopp1976}. CMEs can interact with each other \citep{Lugaz2017}. In the outer corona, the solar wind can accelerate or decelerate the CMEs \citep{Yashiro2004}.}

\edit1{To understand the nature of the CMEs, we need to test these models with experimental data. We need to measure CME acceleration and search for signatures of the acceleration mechanisms.}

\edit1{However, measurements of CME acceleration have difficulties. Today, the early evolution of CMEs is studied with EUV telescopes \citep[for example, AIA, ][]{Lemen2011}, and late evolution with the white-light coronagraphs \citep[for example LASCO/C2, C3; ][]{Brueckner1995}. The gap between these instruments---altitude range of 1.2--2~$R_\odot$---is the ``blind zone'' for the traditional sets of instruments. At the same time, it is in this altitude range the main CME acceleration usually occurs.}

\begin{figure}[hbt]
\centering
\includegraphics[width = 0.45\textwidth]{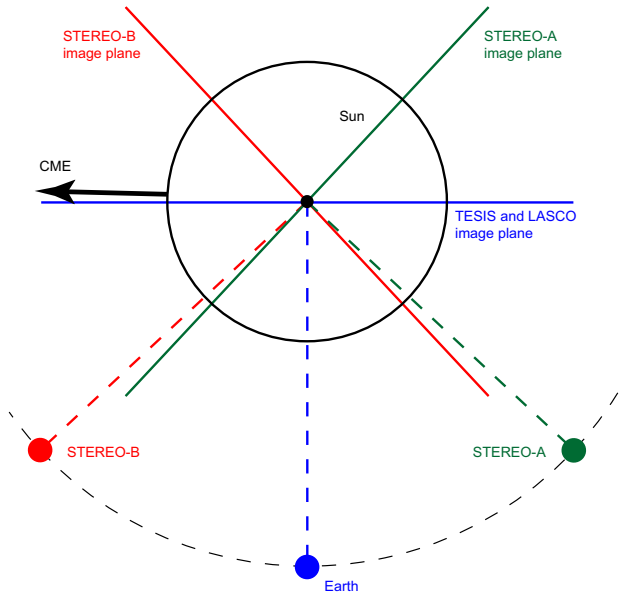}
\caption{\edit2{Position of the STEREO satellites.}}
\label{F:Stereo_position}
\end{figure}

\edit1{Observations of the CMEs in the ``blind zone'' are scarce.} \edit2{They have been performed by the} \edit1{LASCO/C1 coronagraph \citep{Zhang2001},} \edit2{the} \edit1{Siberian Solar Radio Telescope \citep{Alissandrakis2013},} \edit2{the} \edit1{Mauna Loa Observatory \citep{Bemporad2007},} \edit2{the} \edit1{TESIS EUV telescopes \citep{Reva2014, Reva2016b, Reva2016},} \edit2{the} \edit1{SPIRIT EUV coronagraphs \citep{Slemzin2008}, and} \edit2{the} \edit1{SWAP EUV telescope \citep{Byrne2014, Mierla2013, DHuys2017}. To better understand the CME, we need more CME observations in the altitude range of 1.2--2~$R_\odot$ from the} \edit2{Sun's} \edit1{center.}

\edit1{In this work, we study the CME that was observed by the TESIS EUV telescope on April 23, 2009.} \edit2{The} \edit1{TESIS EUV telescope could observe the solar corona  up to distances of 2~$R_\odot$ from the} \edit2{Sun's} \edit1{center. The evolution of this event was observed continuously from the solar surface up to the boundaries of the} \edit2{LASCO} \edit1{field of view.}

\edit1{The studied CME had a complex acceleration profile: two impulsive acceleration phases and one impulsive deceleration phase. The CME showed signatures of several acceleration mechanisms: kink instability, mass drainage, flare reconnection, and a CME-CME collision. The aims of this paper are to present the observations and study the CME acceleration mechanisms.}

\section{Experimental Data}

\begin{table*}[bht]
\centering
\caption{Characteristics of the used instruments.}
\begin{tabular}{llcrrr}
\hline
\hline
Satellite       & Instrument                      & Wavelength  & Resolution            & Field of View       & Cadence    \\
\hline
CORONAS-PHOTON  & TESIS EUV telescope             & 171~\AA     & 3.4$^{\prime\prime}$  & $\le 2$~$R_\odot$   & 30--60~min \\
                &                                 & 304~\AA     & 1.7$^{\prime\prime}$  & $\le 2$~$R_\odot$   & 4~min      \\
                & \ion{Mg}{12} spectroheliograph & 8.42~\AA    & 4.0$^{\prime\prime}$  & $\le 1.6$~$R_\odot$ & 90~min     \\
                & SphinX                          & 1--15~keV   & ---                   & ---                 & 1~min      \\
SOHO            & EIT                             & 304~\AA     & 5.3$^{\prime\prime}$  & $\le 1.6$~$R_\odot$ & 12~min     \\
                & LASCO/C2                        & white light & 11.4$^{\prime\prime}$ & 2--6~$R_\odot$      & 20~min     \\
                & LASCO/C3                        & white light & 56.0$^{\prime\prime}$ & 4--30~$R_\odot$     & 30~min     \\
STEREO          & EUVI                            & 171~\AA     & 1.6$^{\prime\prime}$  & $\le 1.7$~$R_\odot$ & 2.5~min    \\
                &                                 & 195~\AA     & 1.6$^{\prime\prime}$  & $\le 1.7$~$R_\odot$ & 10~min     \\
                &                                 & 284~\AA     & 1.6$^{\prime\prime}$  & $\le 1.7$~$R_\odot$ & 20~min     \\
                &                                 & 304~\AA     & 1.6$^{\prime\prime}$  & $\le 1.7$~$R_\odot$ & 10~min     \\
                & COR1                            & white light & 7.0$^{\prime\prime}$  & 1.5--4~$R_\odot$    & 5--10~min  \\
                & COR2                            & white light & 14.7$^{\prime\prime}$ & 2.5--15~$R_\odot$   & 15~min     \\
\hline
\end{tabular}
\label{T:Instruments}
\end{table*}

In this study, we \edit2{used} the data of the TESIS EUV telescopes and the \ion{Mg}{12} spectroheliograph  \citep{Kuzin2011}, the SphinX spectrophotometer \citep{Gburek2011}, the LASCO coronagraphs \citep{Brueckner1995}, \edit1{the EIT telescope \citep{del95},} and the data of the \textit{STEREO} satellites \citep{Howard2008}.

TESIS is an instrument assembly that observed the solar corona in the soft X-ray and EUV. It worked on board the \textit{CORONAS-PHOTON} satellite \citep{Kotov2011}. The TESIS EUV telescope imaged \edit2{the} solar corona in the Fe~171~\AA\ and He~304~\AA\ lines. The special feature of the TESIS EUV telescope was its ability to image \edit2{the} corona up to a distance of 2~$R_\odot$ from the Sun's center \edit1{in the Fe~171~\AA\ line} \citep[for details, see][]{Reva2014}. We use \edit2{the} TESIS data to observe coronal magnetic structure and prominence evolution at high altitudes.

\edit1{During the period of the observations, the TESIS 304~\AA\ channel observed the Sun with} \edit2{a} \edit1{4~min cadence. However, it had data gaps. To fill these data gaps, we used the data of the EIT 304~\AA\ telescope. During the period of the observations,} \edit2{the} \edit1{EIT telescope observed the Sun in} \edit2{the} \edit1{304~\AA\ line with} \edit2{a} \edit1{12~min cadence.}

The \ion{Mg}{12} spectroheliograph was a part of the TESIS assembly. The instrument built monochromatic images of the solar corona in the \ion{Mg}{12} 8.42~\AA\ line. This line emits at temperatures higher than 4~MK, and the \ion{Mg}{12} images contain only the signal from the hot plasma without any low temperature background.  In 2009, the Sun was in the minimum of its activity cycle, and the spectroheligraph \ion{Mg}{12} registered sub-A class flare-like events \citep{Kirichenko2017a, Kirichenko2017b}. We use the spectroheliograph \ion{Mg}{12} to study the X-ray emission associated with the analyzed CME.

SphinX is a spectrophotometer that worked on board the \textit{CORONAS-PHOTON} satellite. SphinX registered solar spectra in the 1--15~keV energy range. In 2009, the solar cycle was in deep minimum, and the GOES flux usually was below the sensitivity threshold. SphinX is more sensitive than GOES, and we use it to see the variation of the X-ray flux.

LASCO is a set of white-light coronagraphs that observe solar corona from 1.1~$R_\odot$ up to 30~$R_\odot$ (C1, 1.1--3~$R_\odot$; C2, 2--6~$R_\odot$; C3, 4--30~$R_\odot$). In 1998, LASCO C1 stopped working, and today LASCO can only image corona above 2~$R_\odot$. We use LASCO data to study the CME \edit2{evolution} above 2~$R_\odot$.

\textit{STEREO} is a set of two satellites: \textit{STEREO-A}, which moves ahead of the Earth, and \textit{STEREO-B}, which moves behind the Earth. During the period of our observations, \edit2{the} \textit{STEREO} satellites were separated from the Earth by 47$^{\circ}$ (see Figure \ref{F:Stereo_position}). \edit2{The} \textit{STEREO} satellites carry EUVI telescopes that image the solar corona in the 171, 195, 284, and 304~\AA\ lines. We use the \textit{STEREO-B} EUVI data to determine \edit2{the} 3D orientation of the erupting prominence and observe the solar surface that is not seen from the Earth point of view. 

Moreover, \edit2{the} STEREO satellites carry \edit2{two} white-light coronagraphs: COR1, which images corona at distances \edit2{of} 1.3--4~$R_\odot$, and COR2, \edit2{which images corona at distances of} 2--15~$R_\odot$. We use the COR1 and COR2 data from both satellites to enhance the measurements of the CME kinematics. 

\edit1{We listed the characteristics of the instruments in Table~\ref{T:Instruments}.}

\section{Results}

\edit1{We will present the observations in the following way. Firstly, we will outline the observations. Secondly, we will describe the complete sequence of events. Finally, we will present the analysis performed on the observations.}

\subsection{Outline of the Observations}

\begin{figure*}[t]
\centering
\includegraphics[width = \textwidth]{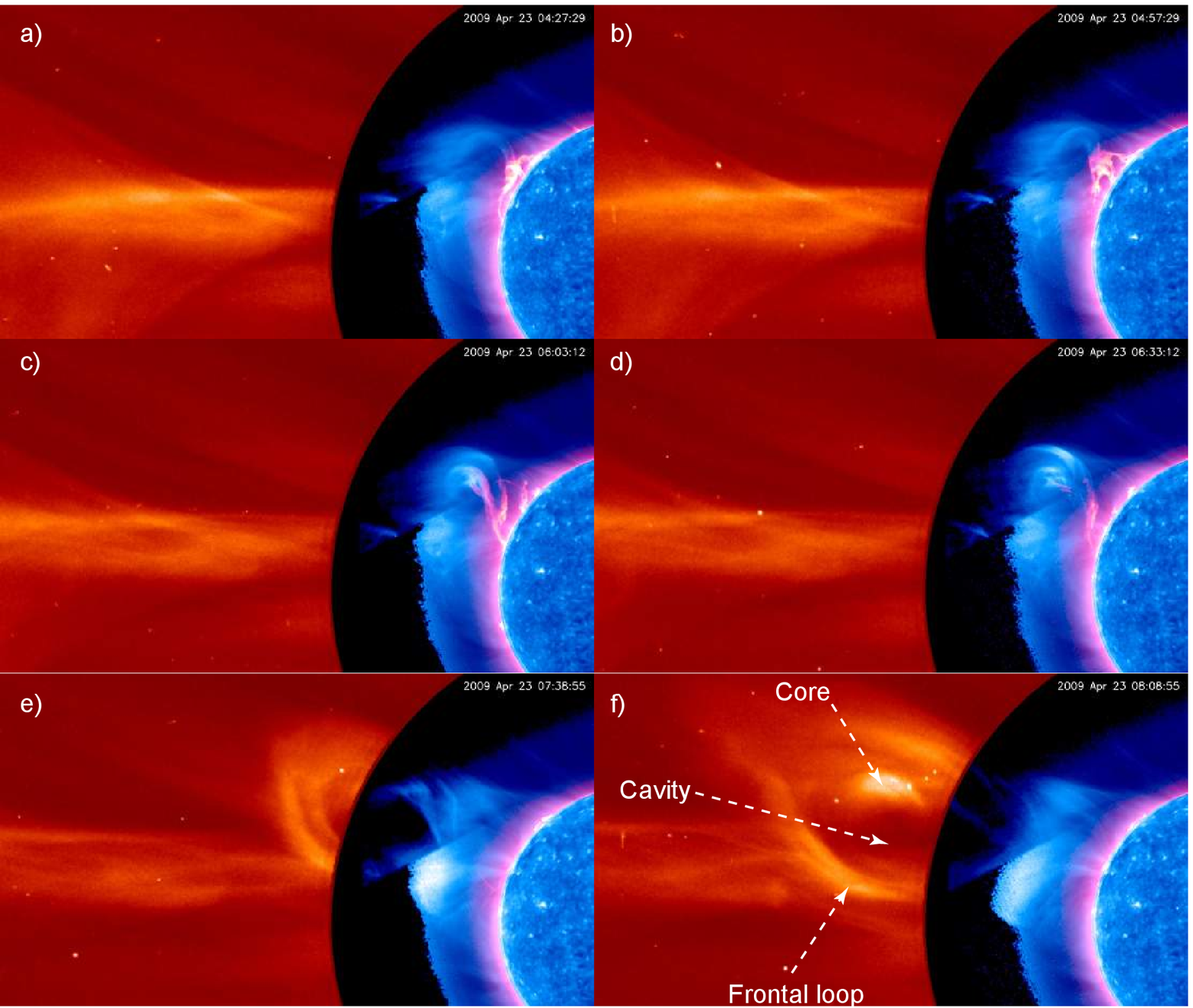}
\caption{\edit2{Early evolution of the studied CME.} Blue: TESIS Fe~171~\AA\ images; purple: TESIS He~304~\AA\ images; red: LASCO/C2 images. An animation is available for this figure.}
\label{F:Fe_He_C2}
\end{figure*}

\edit2{The CME under study} \edit1{occurred at the} \edit2{north-eastern} \edit1{part of} \edit2{the} \edit1{solar limb on April~23, 2009} \edit2{(see Figure~\ref{F:Fe_He_C2}).} \edit1{We recommend the reader to watch the animation to the Figure~\ref{F:Fe_He_C2} before proceeding further.}

\edit1{The studied event evolved in the following way. The prominence started to move up. Later,} \edit2{it} \edit1{twisted in the STEREO-B EUVI images, and a fluxrope appeared in the TESIS Fe~171~\AA\ images.}

\begin{figure*}[t]
\centering
\includegraphics[width = 0.95\textwidth]{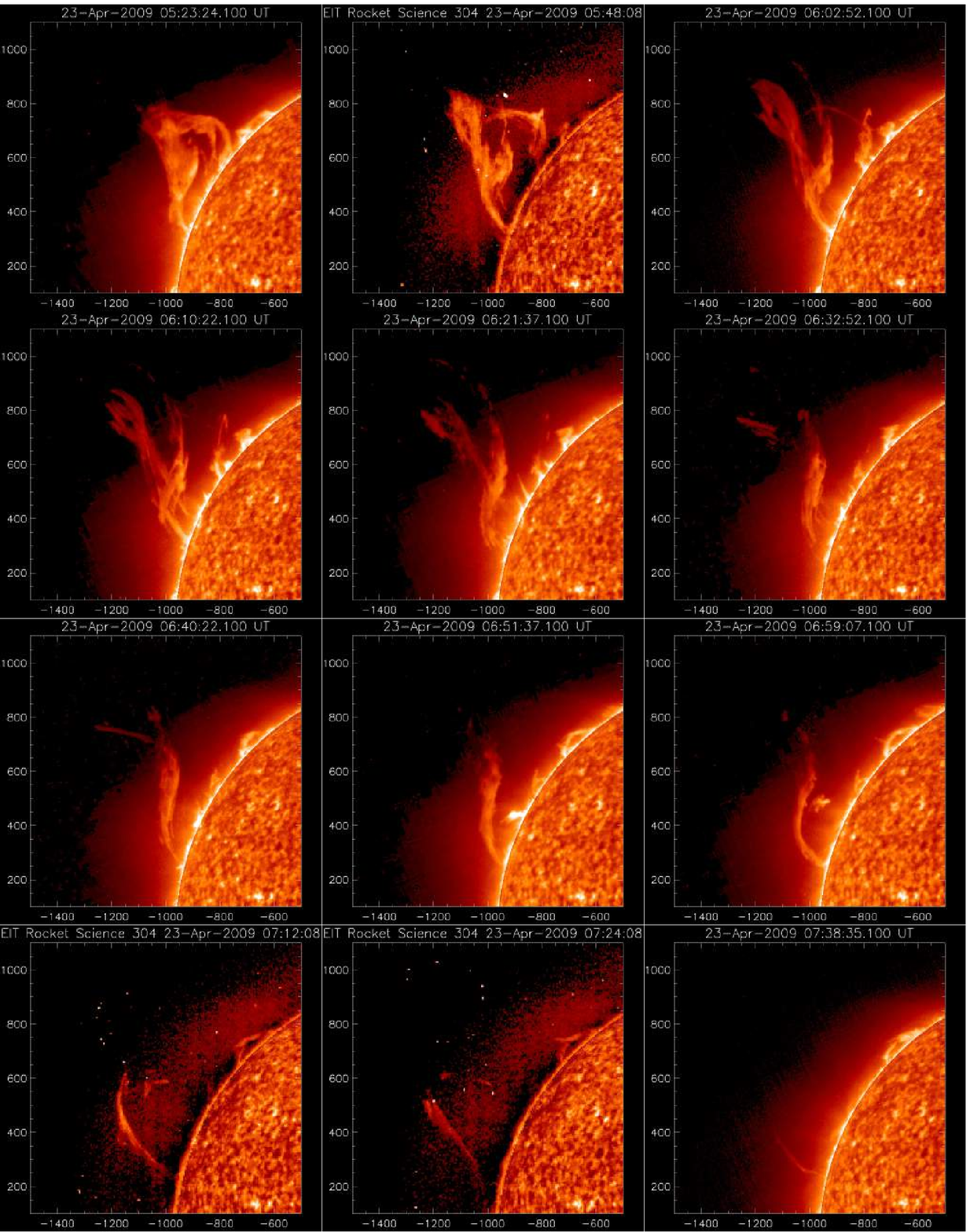}
\caption{Draining of the prominence. These are TESIS and EIT  He~304~\AA\ images. \edit2{To increase the prominence's visibility, we lowered the disk intensity (multiplied by 0.2).} Coordinates are measured in arcseconds. An animation is available for this figure.}
\label{F:Prominence_drain}
\end{figure*}

\edit1{At the altitude of 220~Mm} \edit2{(0.32~$R_\odot$)} \edit1{above the solar surface,} the prominence tore apart and drained down, while the fluxrope continued to erupt \edit1{(see Figures~\ref{F:Fe_He_C2}} \edit2{and} \edit1{\ref{F:Prominence_drain})}. In the LASCO/C2 images, the fluxrope corresponded to the CME core. By the time the fluxrope had reached the LASCO/C2 field of view, the prominence had already drained down to the solar surface.

\edit1{At the eastern solar limb, another CME occurred before the studied one. In the LASCO/C2 field of view, the studied CME collided with the preceding CME.}

\edit2 {The kinematics of the studied CME} \edit1{had} \edit2{an} \edit1{impulsive acceleration phase,} \edit2{an} \edit1{impulsive deceleration phase, and a second impulsive acceleration phase}. During the first acceleration phase, we observed signatures of flare reconnection: an increase in the X-ray flux and a two-ribbon flare. There were no signatures of \edit1{flare} reconnection during the second acceleration phase. \edit2{The second acceleration and the deceleration phases} \edit1{occurred after the CME-CME collision.}

\begin{figure*}[t]
\centering
\includegraphics[width = 0.75\textwidth]{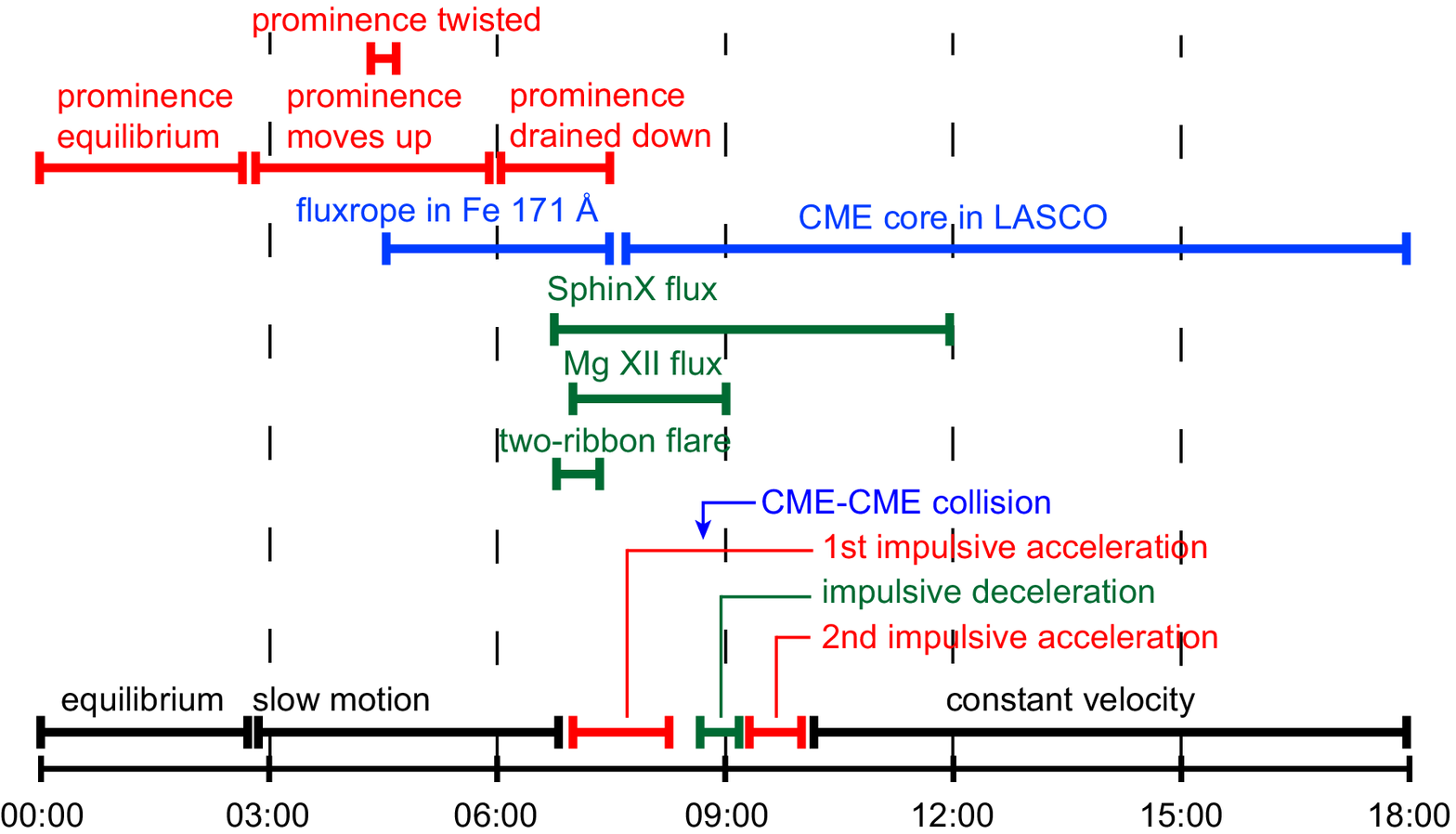}
\caption{Summary of the observations.}
\label{F:Summary}
\end{figure*}

\edit1{The observations are summarized in Figure~\ref{F:Summary} and Table~\ref{T:Summary}.}

\begin{table}[bht]
\centering
\caption{Summary of the observations.}
\begin{tabular}{p{0.28\linewidth}p{0.64\linewidth}}
\hline
\hline
Time, UT                      & Description                                \\
\hline
$\approx$~02:51        & prominence started to move up                     \\
$\approx$~04:27        & CME core appeared in the TESIS Fe~171~\AA\ images \\
\edit1{$\approx$~04:35}& \edit1{prominence twisted}                        \\
$\approx$~06:02        & prominence started to drain                       \\
$\approx$~07:38        & prominence drained down                           \\
$\approx$~07:38        & the core passed to the LASCO/C2 field of view     \\
\edit1{$\approx$~07:00--08:15} & the first impulsive acceleration phase    \\
$\approx$~06:45--12:00 & the SphinX  X-ray flux                            \\
$\approx$~07:00--09:00 & the \ion{Mg}{12} spectroheliograph X-ray flux     \\
$\approx$~06:46--07:20 & motion of the flare ribbons                       \\
\edit1{$\approx$~08:50} & \edit1{CME-CME collision}                        \\
\edit1{$\approx$~08:40--09:15} & \edit1{impulsive deceleration phase}      \\
\edit1{$\approx$~09:15--10:00} & the second impulsive acceleration phase   \\
\hline
\end{tabular}
\label{T:Summary}
\end{table}

\subsection{Sequence of Events}

\begin{figure*}[hbt]
\centering
\includegraphics[width = 0.9\textwidth]{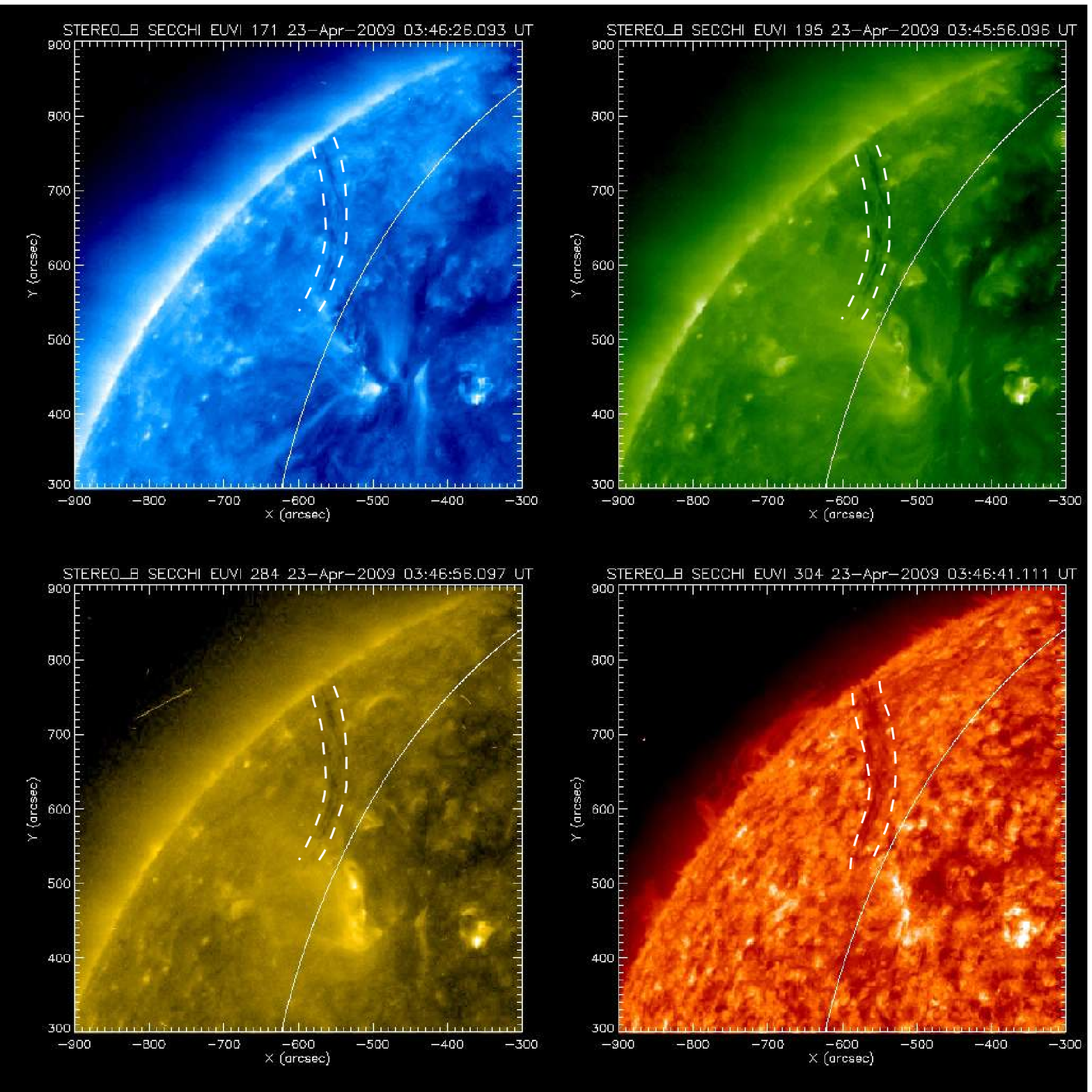}
\caption{Erupting prominence (filament) in the STEREO EUVI-B images. Dashed lines mark the filament; \edit2{the} solid line designates the TESIS limb.}
\label{F:STEREO_EUVI_B}
\end{figure*}

\edit1{The studied CME was associated with a prominence.} \edit2{The} \edit1{TESIS He~304~\AA\ images} \edit2{indicate that} \edit1{the prominence was located at the the} \edit2{north-eastern} \edit1{part of the solar limb.} In the STEREO-B EUVI images, the prominence looked like a filament \edit2{that} was inclined by 25$^{\circ}$ to the meridian and was close to the TESIS limb (see Figure~\ref{F:STEREO_EUVI_B}). \edit2{Therefore,} the prominence eruption occurred almost entirely in the TESIS image plane (see Figure~\ref{F:Stereo_position}).

\begin{figure}[hbt]
\centering
\includegraphics[width = 0.45\textwidth]{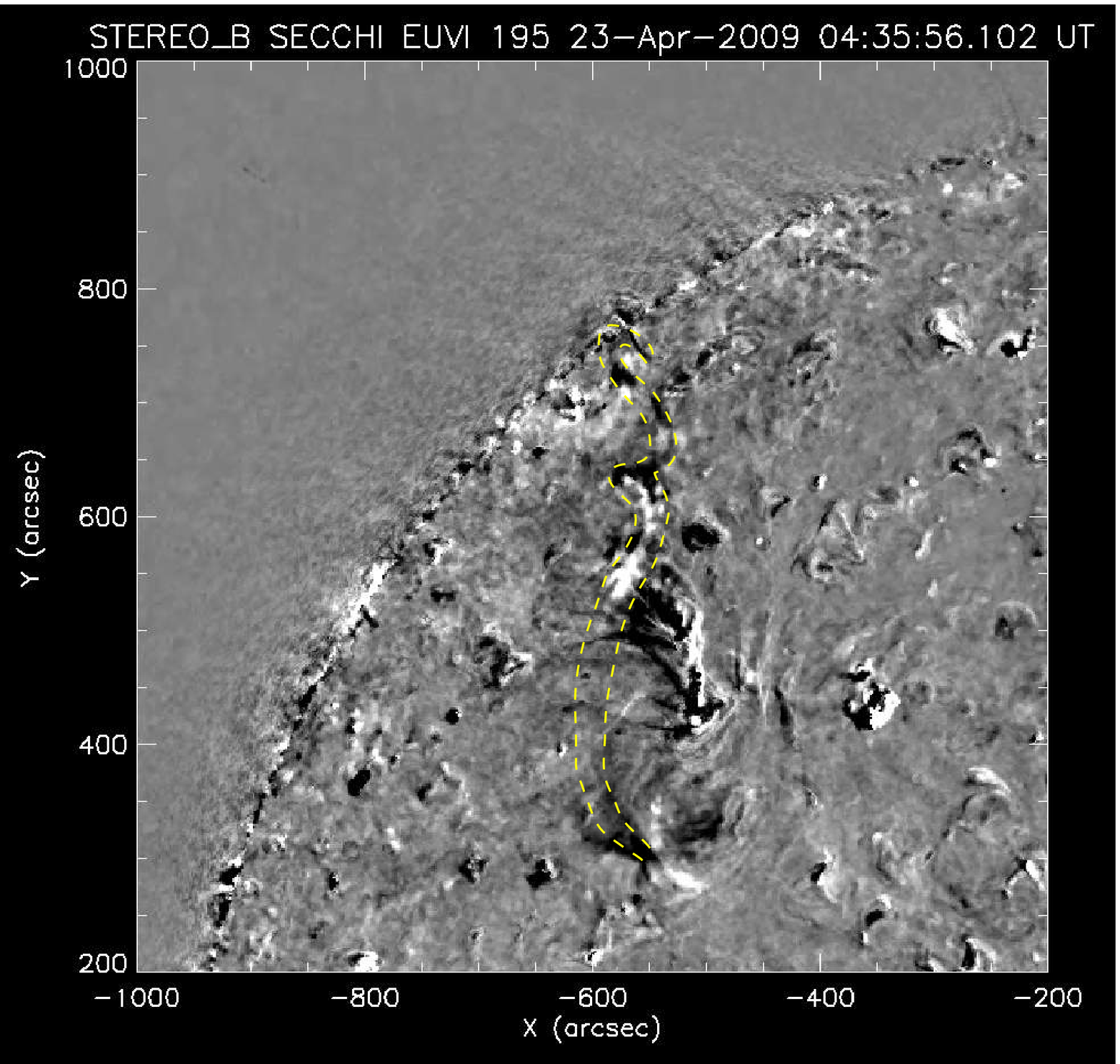}
\caption{Twisting of the prominence. This is a running difference image \edit2{from} the EUVI-B 195 \AA\ channel. \edit2{The yellow} dashed lines indicate the prominence.}
\label{F:twisting}
\end{figure}

The prominence started to lift at 02:51~UT on \edit2{April~23, 2009}. At 04:28~UT, in the Fe~171~\AA\ images, a fluxrope formed. \edit1{In the STEREO-B EUVI images,} the prominence twisted at $\approx$~04:35~UT (see Figure~\ref{F:twisting}).

\begin{figure*}[hbt]
\centering
\includegraphics[width = 0.95\textwidth]{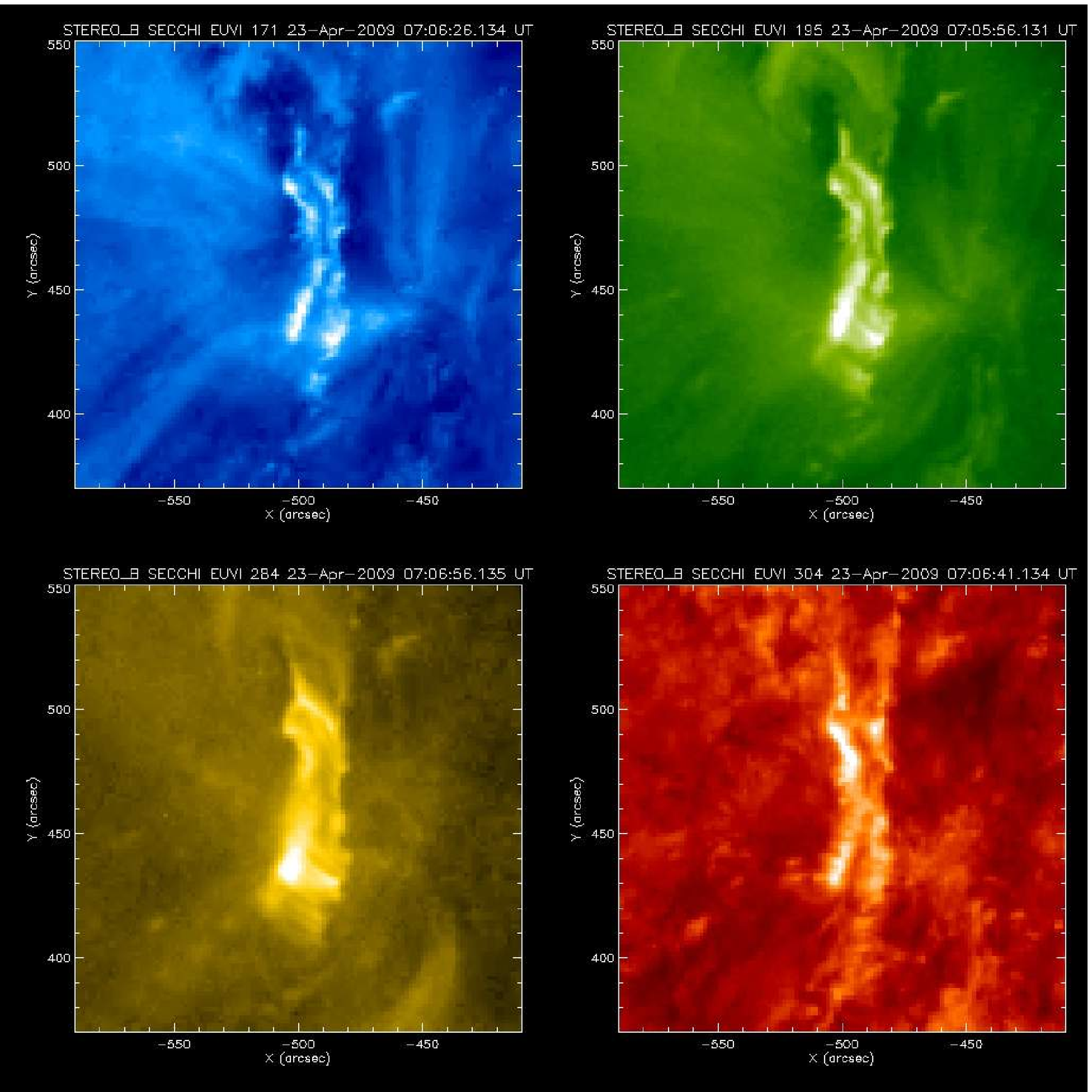}
\caption{Two-ribbon flare observed by STEREO-B EUVI.}
\label{F:two_ribbon_flare}
\end{figure*}

\edit2{From 06:46 to 07:20~UT, the flare ribbons appeared below the prominence in all four channels of the STEREO-B EUVI (see Figure~\ref{F:two_ribbon_flare}).}

\begin{figure}[t]
\centering
\includegraphics[width = 0.47\textwidth]{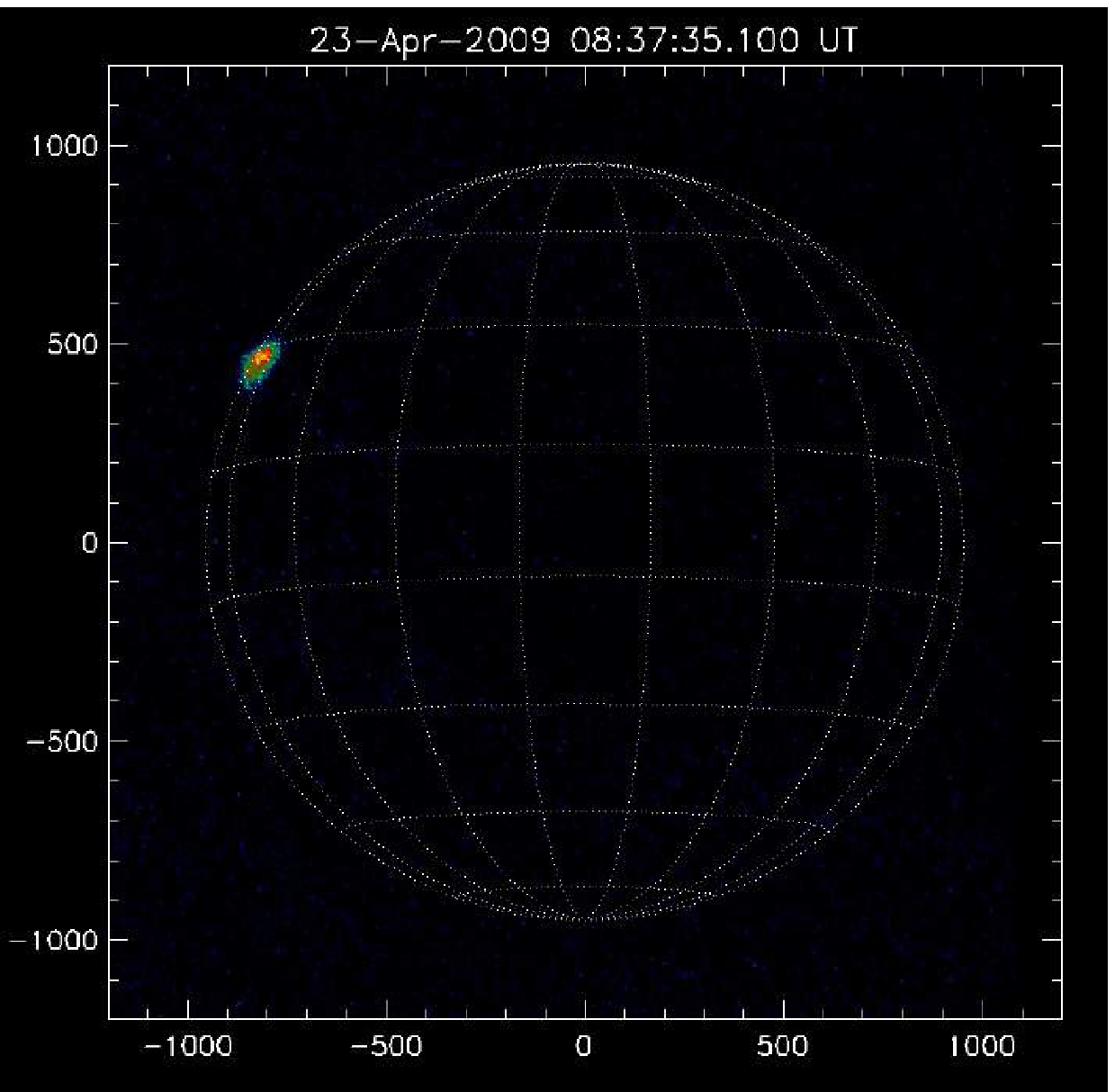}
\caption{\edit2{Hot plasma observed below the studied CME with the \ion{Mg}{12} spectroheliograph. Yellow and red correspond to high intensities, while blue and green correspond to low intensities.} Coordinates are measured in arc seconds.}
\label{F:mg12_image}
\end{figure}

We observed \edit2{an X-source} in the \ion{Mg}{12} spectroheliograph  images \edit1{from 07:00 to 09:00~UT} (see Figure~\ref{F:mg12_image}). This source corresponds to the flare arcade below the CME. It was the only X-ray source on the Sun. 

\begin{figure*}[t]
\centering
\includegraphics[width = \textwidth]{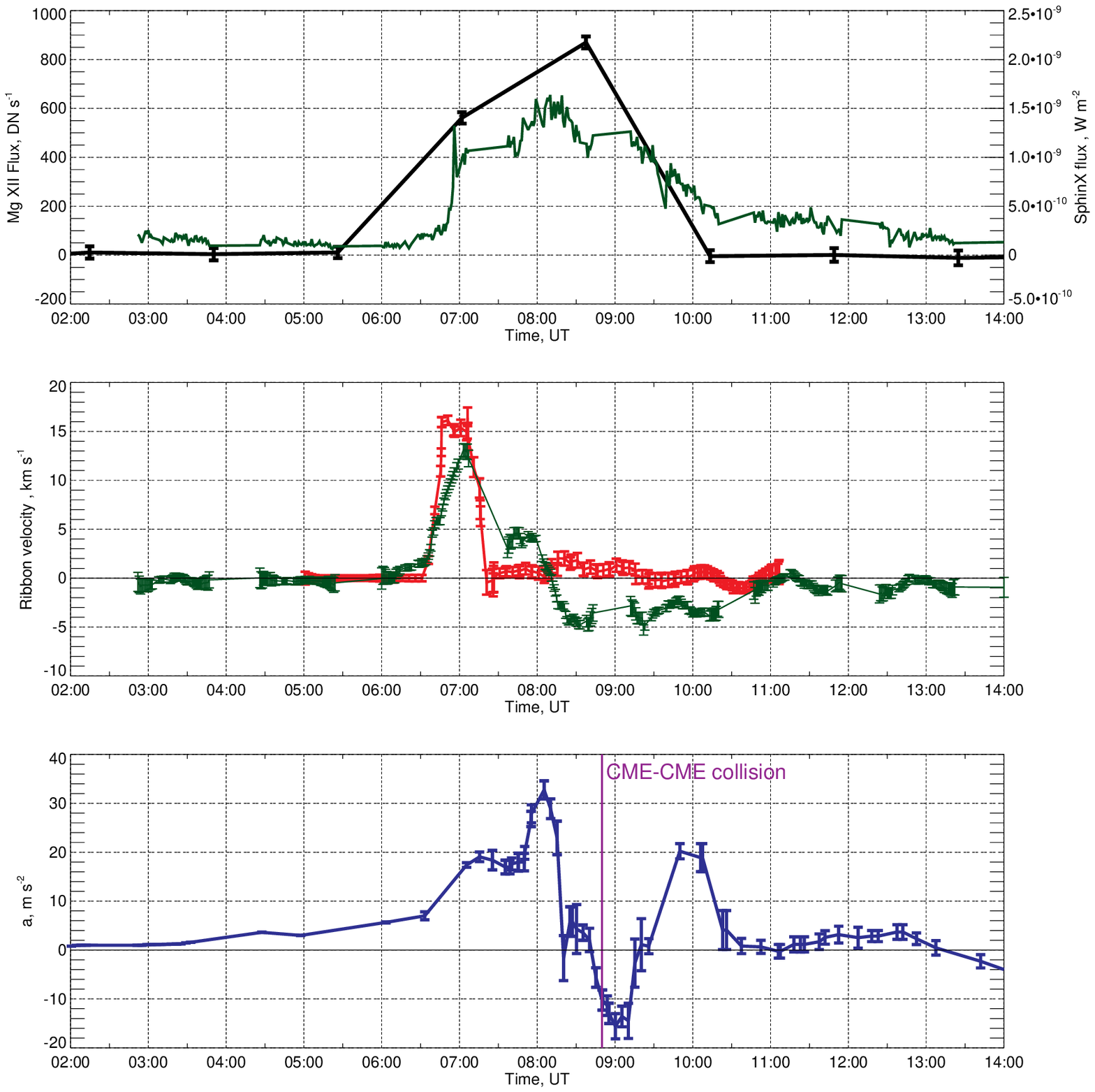}
\caption{\edit2{Comparison of the X-ray flux, the acceleration of the CME core, and the separation velocity of the flare ribbons.} Black: \edit2{the flux of the hot source observed with the \ion{Mg}{12} spectroheliograph}; green top: the SphinX flux convolved with the response function of the GOES 1--8~\AA\ channel; green \edit3{middle}: the \edit2{derivative of the} SphinX flux; red: the \edit2{separation} velocity of the flare \edit2{ribbons}; blue: \edit2{the acceleration of the CME core}. \edit2{Purple line marks the time of the CME-CME collision.}}
\label{F:mg_flux}
\end{figure*}

During the period of the observations, the GOES X-ray flux \edit2{was below its} sensitivity threshold. \edit2{From $\approx$~06:45~UT to 11:00~UT}, the SphinX registered an increase of the  X-ray flux. Since the \ion{Mg}{12} spectroheliograph observed only one X-ray source, the increase of the SphinX flux comes entirely from the flare arcade. 

We convolved the SphinX spectrum with the response function of the GOES 1--8~\AA\ channel and obtained a synthetic GOES flux (see Figure~\ref{F:mg_flux}, top). The studied flare arcade \edit2{was of the} A0.15 GOES class.

The prominence was located in the lowest parts of the magnetic structure visible in the Fe~171~\AA\ images. While erupting, the magnetic structure dragged the prominence up. At the altitude of 220~Mm \edit2{(0.32~$R_\odot$)} above the solar surface, the prominence tore apart and started to drain while the fluxrope continued to erupt \edit1{(see Figures~\ref{F:Fe_He_C2}} \edit2{and} \edit1{\ref{F:Prominence_drain})}.

The gap between the LASCO/C2 and TESIS fields of view is small. This allows us to track the CME without losing it from sight. In the LASCO/C2 images, the CME had a three-part structure: core, cavity, and frontal loop (see Figure~\ref{F:Fe_He_C2}f). The CME core in the LASCO images corresponds to the fluxrope center observed in the Fe~171~\AA\ images. When the CME core \edit2{left} the TESIS  field of view, the prominence had already drained down.

\begin{figure*}[t]
\centering
\includegraphics[width = \textwidth]{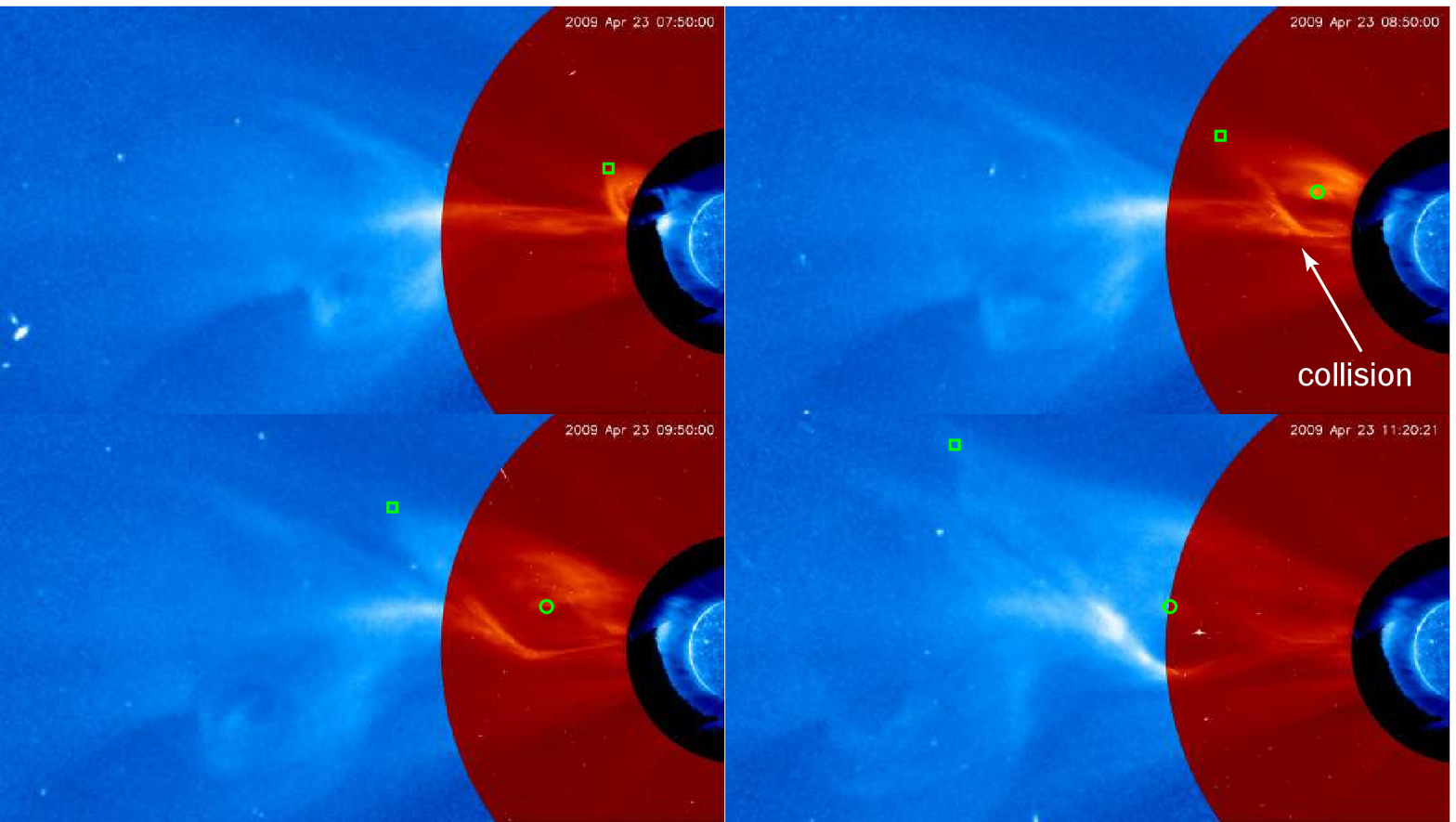}
\caption{CME-CME collision. Outer blue: LASCO/C3 images, red: LASCO/C2 images, inner blue: TESIS Fe~171~\AA\ images. Circle marks the CME core. Rectangle marks part of the frontal loop. Marked parts were used to measure CME kinematics. An animation is available for this figure.}
\label{F:CME_CME_collision}
\end{figure*}

\edit1{At the eastern solar limb, another CME occurred before the studied one. In the LASCO/C2 field of view, the studied CME collided with the preceding CME at $\approx$~08:50~UT (see Figure~\ref{F:CME_CME_collision}).}

\subsection{Analysis}

\subsubsection{CME Kinematics}

\begin{figure*}[t]
\centering
\includegraphics[width = 0.89\textwidth]{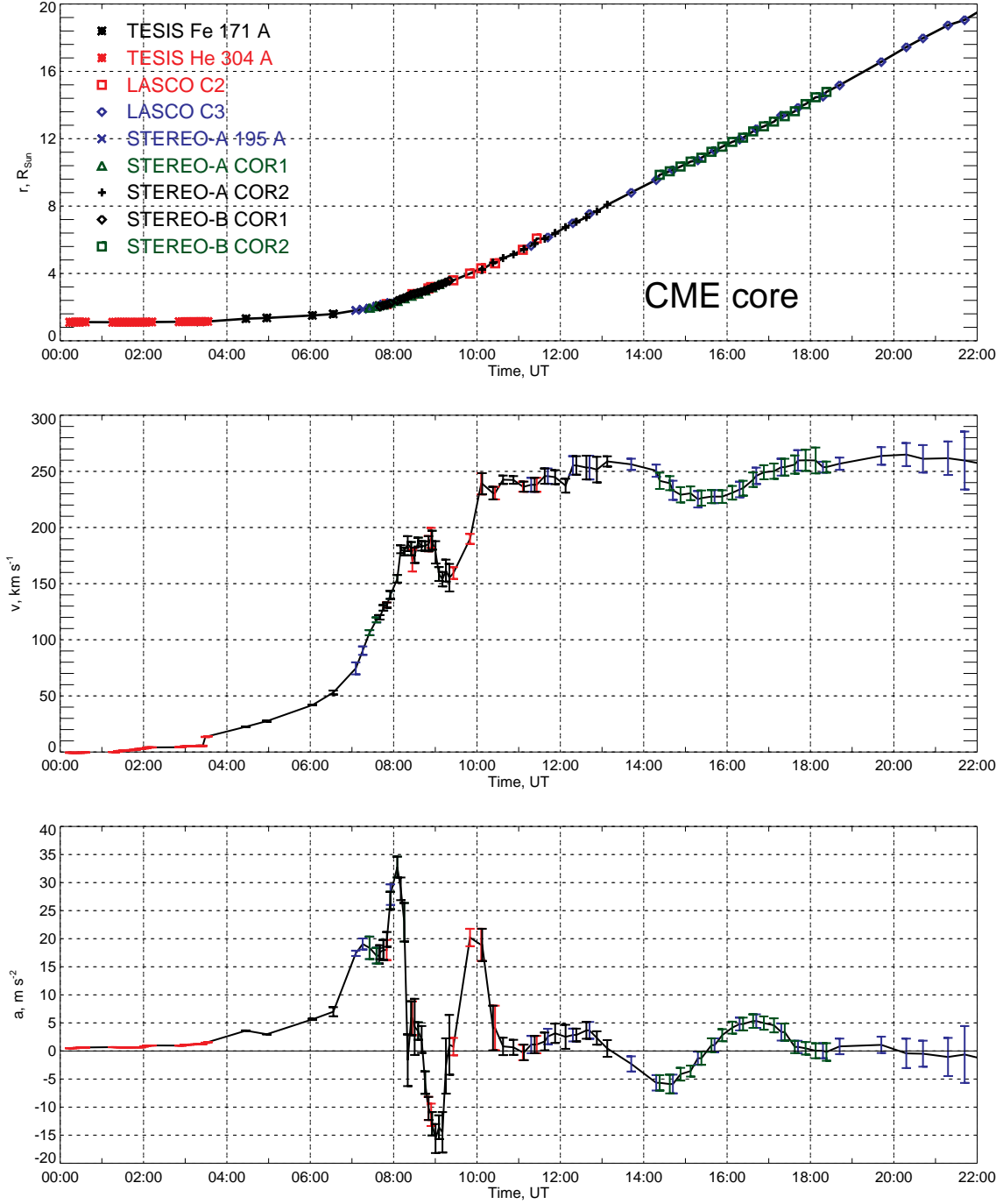}
\caption{\edit2{The kinematics of the CME core}. $r(t)$ is the distance from the CME core to the \edit2{Sun's} center, $v(t)$ is the CME core radial velocity, \edit2{and} $a(t)$ is the CME core radial acceleration.}
\label{F:Kinematics_core}
\end{figure*}

\begin{figure*}[t]
\centering
\includegraphics[width = 0.90\textwidth]{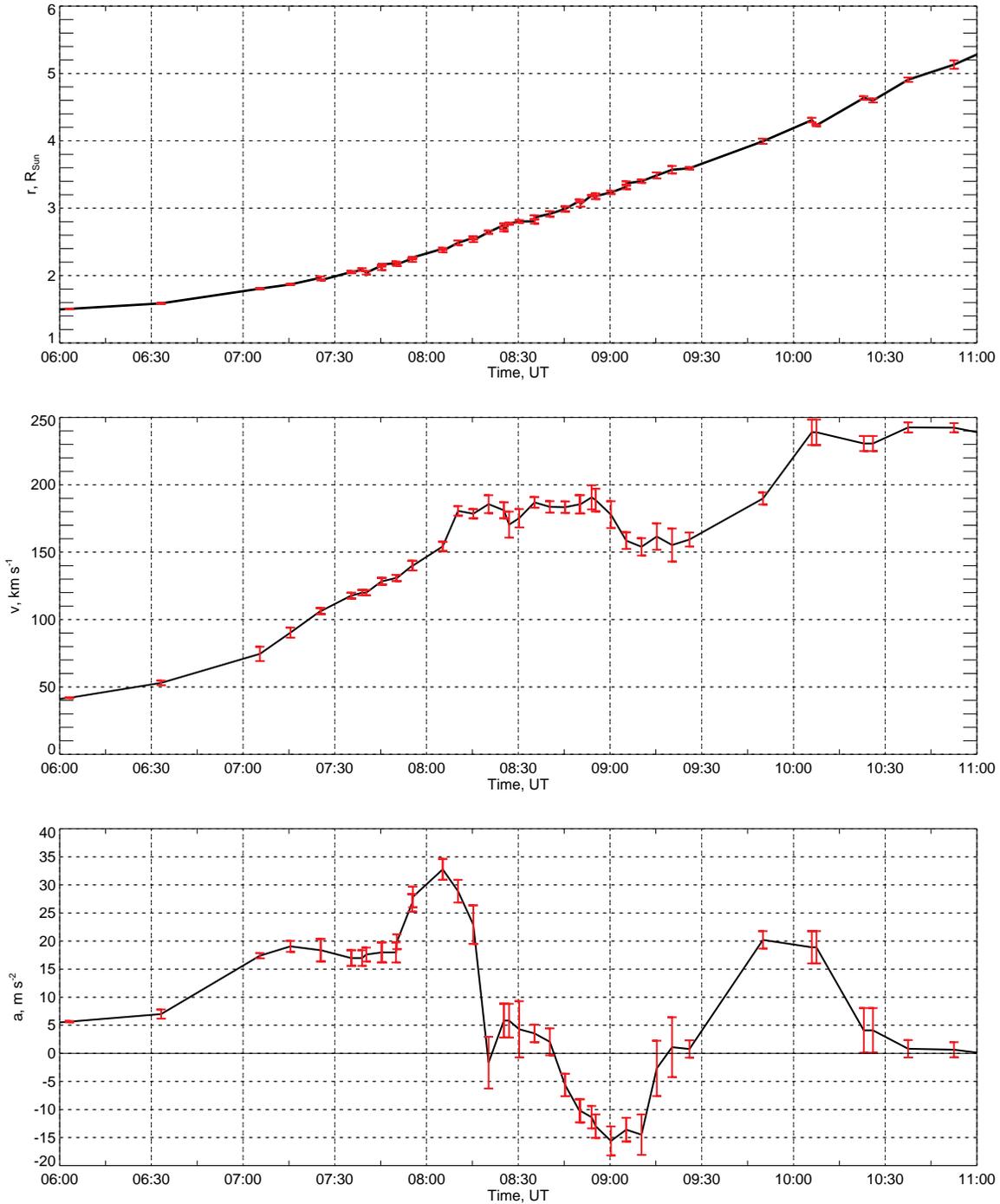}
\caption{\edit2{The kinematics of the CME core. Zoomed version of Figure~\ref{F:Kinematics_core}}. $r(t)$ is the distance from the CME core to the \edit2{Sun's} center, $v(t)$ is the CME core radial velocity, \edit2{and} $a(t)$ is the CME core radial acceleration.}
\label{F:Kinematics_core_zoomed}
\end{figure*}

\begin{figure*}[t]
\centering
\includegraphics[width = 0.90\textwidth]{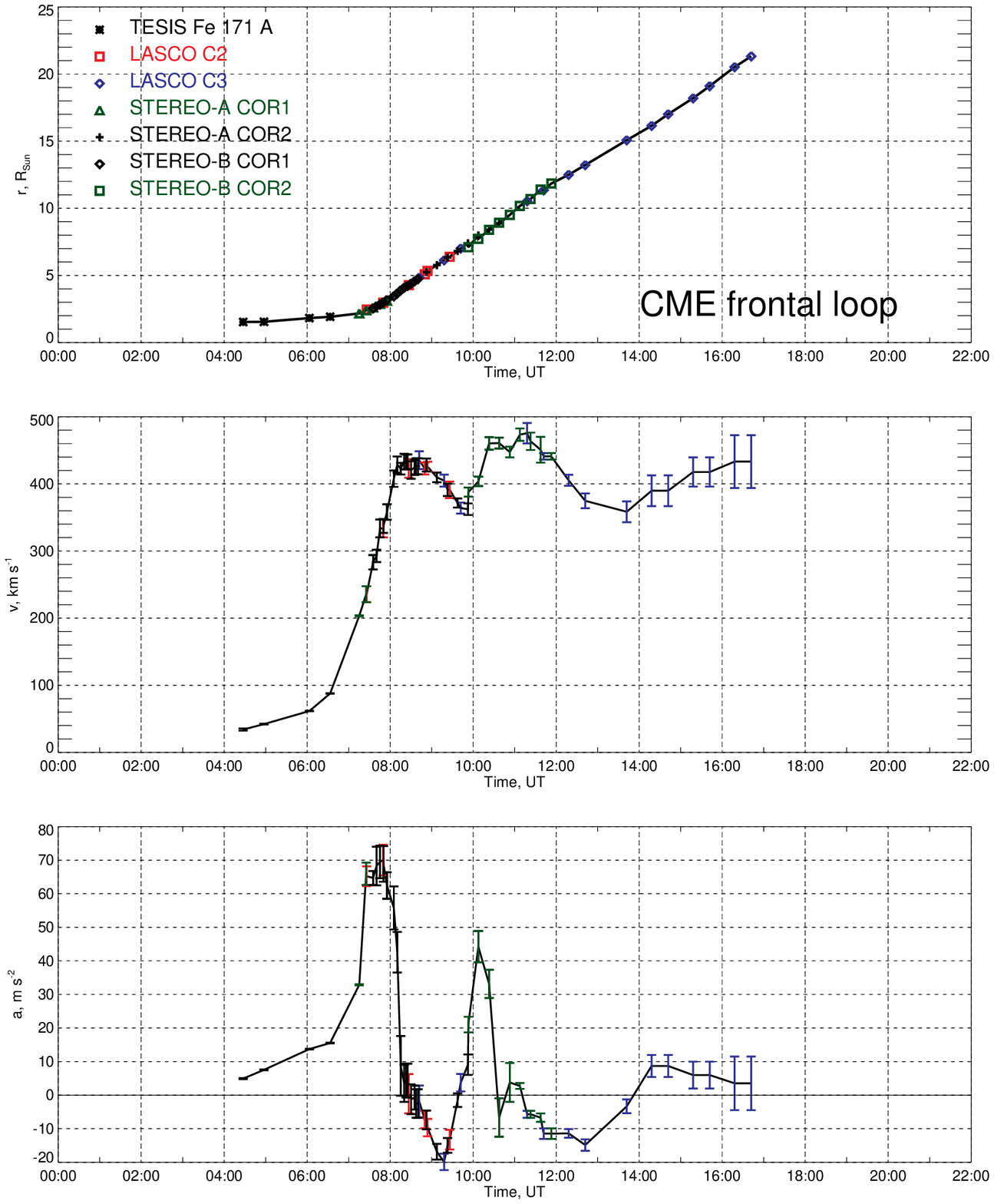}
\caption{\edit2{Kinematics of the CME frontal loop.} $r(t)$ is the distance from the CME frontal loop to the \edit2{Sun's} center, $v(t)$ is the CME frontal loop radial velocity, \edit2{and} $a(t)$ is the CME frontal loop radial acceleration.}
\label{F:Kinematics_frontal_loop}
\end{figure*}

We measured the coordinates of \edit1{the CME core (its central part) and the CME frontal loop (its farthest part)} in the TESIS Fe 171~\AA\ and He~304~\AA, LASCO C2 and C3 images \edit2{(see Figure~\ref{F:CME_CME_collision}).} \edit1{We used} a simple point and click method. \edit1{The procedure was repeated} \edit2{nine} times to make this method less subjective, and \edit2{to} estimate error bars. \edit2{Using the obtained coordinates, we calculated the distance from the Sun's center ($r(t)$).}

The studied CME was also observed by the \textit{STEREO-A} (EUVI 195~\AA, COR1 and COR2) and \textit{STEREO-B} (COR1 and COR2). We measured the CME core \edit1{and frontal loop} trajectories in these images using the same method.

\edit2{We want to combine the kinematics measured with TESIS, LASCO, and STEREO. However,} TESIS/LASCO, \textit{STEREO-A}, and \textit{STEREO-B} observed the event from different view-points. The instruments responded differently to temperature and density. The positions of the CME core \edit1{and frontal loop were} determined subjectively. Due to these factors, if we put the values $r(t)$ measured by different instruments on the same plot, the values will slightly differ from each other (even after the correction of the projection effects).

To correct the discrepancy, we adopted the procedure from \citet{Reva2016}. First, we scaled the values obtained from \textit{STEREO-A} and \textit{STEREO-B} using the separation angles of the satellites. After this procedure, the \textit{STEREO} points slightly deviated from the TESIS/LASCO points. 

We assumed that the CME expanded uniformly, and scaled each of the \textit{STEREO} channels to fit the TESIS/LASCO plot (see Table~\ref{T:Multiplier}). As a result, we obtained a composite kinematics plot consisting of 139~points \edit1{for the CME core and 61~points for the frontal loop (see Figures~\ref{F:Kinematics_core}, \ref{F:Kinematics_core_zoomed}, and \ref{F:Kinematics_frontal_loop}, top).}

\begin{table}[bht]
\centering
\caption{Kinematics scaling multiplier.}
\begin{tabular}{lll}
\hline
\hline
Channel                        & Core & Frontal loop \\
\hline
\textit{STEREO-A} EUVI 195~\AA & 0.93 & ---          \\
\textit{STEREO-A} COR1         & 0.83 & 0.82         \\
\textit{STEREO-A} COR2         & 1.00 & 0.80         \\
\textit{STEREO-B} COR1         & 0.82 & 0.90         \\
\textit{STEREO-B} COR2         & 0.97 & 0.84         \\
\hline
\end{tabular}
\label{T:Multiplier}
\end{table}

We \edit2{want to} numerically differentiate $r(t)$ and obtain radial velocity $v(t)$ and radial acceleration $a(t)$. \edit2{Since the studied CME collided with another one, we expect its acceleration profile to be complex. Therefore, we should use a numerical differentiation method that does not make any assumptions about the shape of the acceleration profile.}

\edit2{Numerical differentiation of the experimental data is a complex task, because the measurements are prone to errors. Application of the straightforward finite difference method will lead to a highly noisy result. To tackle this problem, a lot of numerical differentiation techniques were developed \citep[for a review, see ][]{Wood1982}.}

\edit1{To find derivatives, we} \edit2{chose the} \edit1{local least-square approximation} \edit2{ method \citep[see pages~320--324  of ][]{Wood1982}.} \edit1{For each point, we} \edit2{fitted} \edit1{the plot with a line using only data from the small vicinity of the point.} \edit2{The tangent of the slope of the fitted line is the derivative.}

\edit1{The size of the vicinity affects the result. If the vicinity is too large, the derivative will be smooth, but the fast-changing details of the derivative will be lost. If the vicinity is too small, we will see the fast dynamics of the derivative, but the derivative will be too} \edit2{noisy.} 

\edit1{In our research, we varied the size of} \edit2{the} \edit1{vicinity along the plot. We wanted to make the size of the vicinity as small as possible, and, at the same time, keep the derivative smooth.}

\edit1{The result is presented in Figures~\ref{F:Kinematics_core}, \ref{F:Kinematics_core_zoomed}, and \ref{F:Kinematics_frontal_loop}. The kinematics of the CME core and frontal loop look similar.}

\begin{enumerate}
\item \edit1{Before $\approx$~07:00~UT, the CME slowly accelerated.}
\item \edit1{From $\approx$~07:00 to $\approx$~08:15~UT, the CME impulsively accelerated.}
\item \edit1{From $\approx$~08:15 to $\approx$~08:40~UT, the CME moved with a constant velocity.}
\item \edit1{From $\approx$~08:40 to $\approx$~09:15~UT, the CME impulsively decelerated.}
\item \edit1{From $\approx$~09:15 to $\approx$~10:00~UT, the CME impulsively accelerated.}
\item \edit1{After $\approx$~10:00~UT, the CME moved with approximately constant velocity.}
\end{enumerate}

\subsubsection{Flare Ribbons Motion}

During the eruption, \edit2{a} two-ribbon flare occurred below the CME (see Figure~\ref{F:two_ribbon_flare}). The speed of the ribbon separation reflects the reconnection rate in the current sheet \citep{Qiu2004}. The faster the ribbons move, the higher the reconnection rate.

We measured the distance between the ribbons in the EUVI-B images using a point-and-click procedure. The main source of errors in this method is the subjectivity. To eliminate subjectivity and estimate error bars, we repeated the procedure multiple times. \edit2{Then} we corrected the result for the projection effect.

Before 06:46~UT, the two-ribbon flare was not seen in the EUVI-B images. We added ``zero points'' to the plot: zero values at times when the EUVI-B took images, but the two-ribbon flare was absent.

\begin{figure}[t]
\centering
\includegraphics[width = 0.49\textwidth]{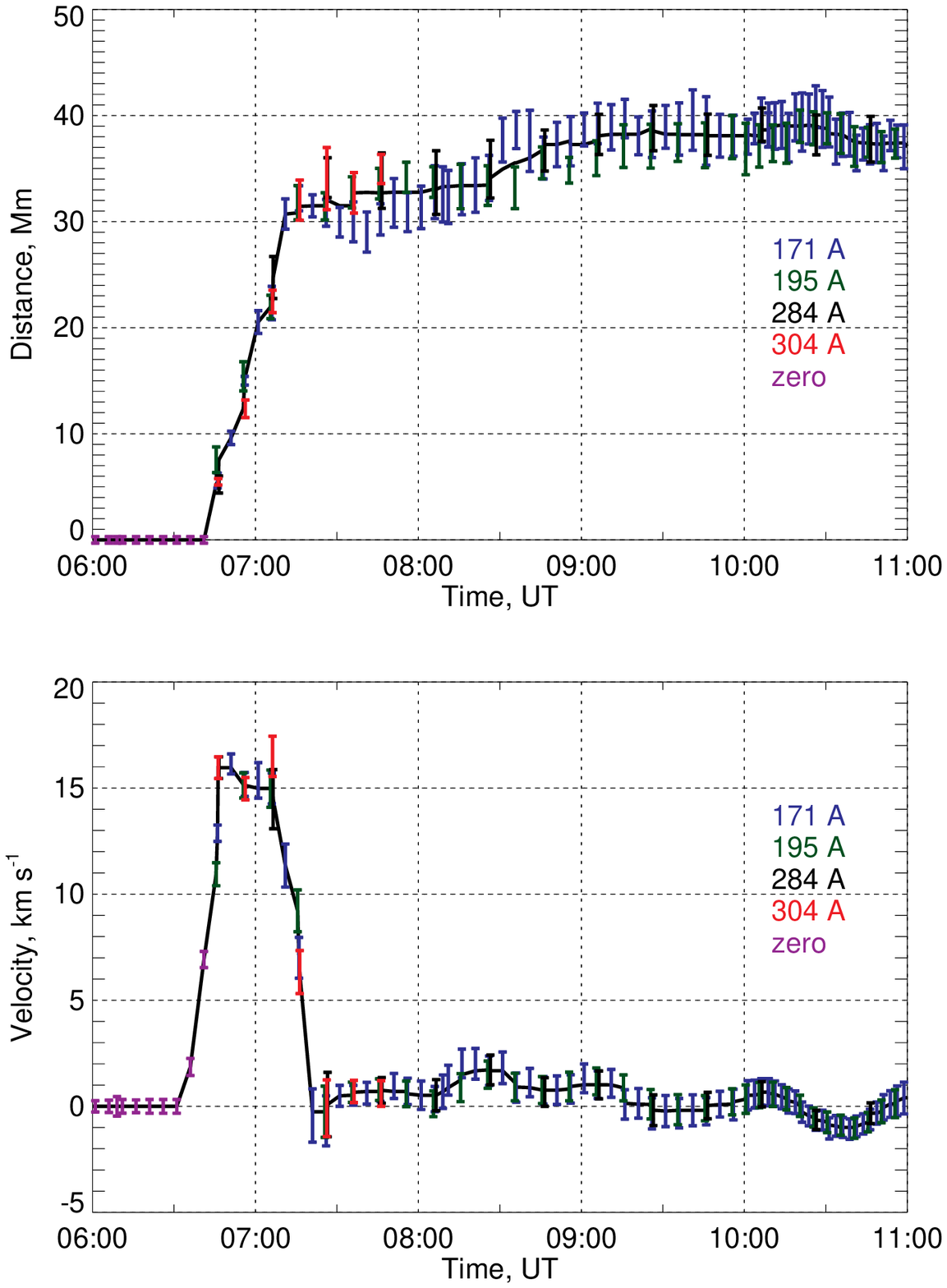}
\caption{\edit2{The distance between the flare ribbons (top) and the velocity of the ribbon separation (bottom).} Blue: EUVI-B 171~\AA\ channel; green: EUVI-B 195~\AA\ channel; black: EUVI-B 284~\AA\ channel; red: EUVI-B 304~\AA\ channel; purple: ``zero points'' (171~\AA\ images without flare ribbons).}
\label{F:arcade_footpoint}
\end{figure}

\edit2{Using local least-square approximation method, we numerically differentiated  the measured distances and obtained the ribbons' separation velocity.} From 06:45~UT to 07:15~UT, the ribbons moved with a roughly constant speed of 15~km~s$^{-1}$ (see Figure~\ref{F:arcade_footpoint}). After 07:15~UT, the distance between the footpoints was constant within error margins. 

\subsubsection{X-ray Flux}

The derivative of the soft X-ray flux reflects the reconnection rate \citep{Neupert1968}. \edit2{Using the local least-square approximation method, we numerically differentiated  the SphinX X-ray flux.}

The derivative of the SphinX flux and the flare \edit2{ribbons} velocity have peaks that correlate with the first impulsive CME acceleration phase (see Figure~\ref{F:mg_flux}, bottom). During the second \edit2{acceleration} phase, there were no \edit2{peaks of the SphinX X-ray flux or ribbon velocity}.

\section{Discussion}

\subsection{Initiation of the CME}

\edit1{In the early stages, the prominence associated with the CME twisted (see Figure~\ref{F:twisting}). We think that this is a manifestation of helical kink instability \citep{Hood1981}. Later, the prominence drained down. Below, we will discuss the role of these two mechanisms in the triggering of the observed CME.}

\edit1{The kink instability could lead to a CME or a failed eruption. \citet{Torok2005} showed that, if the magnetic field falls off rapidly in the vicinity of the fluxrope, then kink instability will lead to the eruption. Otherwise, a failed eruption will happen.}

In \edit2{the} \edit1{mass-unload model}, a CME erupts due to the draining of mass from the CME pre-erupting structure \citep{Fan2003}. \edit1{If the magnetic field is strong, the effect of gravity is negligible. If the magnetic field is weak, the presence of mass will make the fluxrope more stable \citep{Reeves2005}.}
If, for some reason, part of the CME mass drains out, \edit1{the equilibrium could be lost}, and the CME will erupt.

\edit2{Our observations differ from the mass-unload model.} \edit1{In the mass-unload model and observations that support it \citep{Seaton2011}}, the CME loses mass before an eruption. In our observations, the CME lost the mass \edit1{after the onset of} the eruption. 

\edit1{\citet{Aulanier2014} analyzed existing CME models and came to the conclusion that only two physical mechanisms can drive the onset of} \edit2{a} \edit1{CME: torus instability \citep{Kliem2006} and breakout reconnection \citep{Antiochos1999}. Other mechanisms could facilitate} \edit2{an} \edit1{eruption or bring the system to the point where torus instability or breakout reconnection will occur.}

\edit1{We see two possible ways to interpret the observations. The first one is that, after} \edit2{experiencing} \edit1{kink instability, the fluxrope reached} \edit2{an} \edit1{altitude where the magnetic field} \edit2{fell} \edit1{off rapidly enough for the torus instability to occur. In this scenario, the torus instability drove the CME onset.} \edit2{The} \edit1{draining played} \edit2{only} \edit1{an auxiliary role: it decreased the CME mass, which helped to increase the CME acceleration.}

\edit1{The second option is that, after the kink instability, the fluxrope reached a meta-stable state. In this case, the draining caused CME onset.}

\edit1{However, we observe that the prominence was continuously rising without reaching any meta-stable state. Therefore, we think that the first interpretation explains the observations more correctly than the second one.}

\subsection{Two Impulsive Acceleration Phases}

The studied CME had two impulsive acceleration phases. \edit1{A similar effect was reported by \citet{Su2012} and \citet{Byrne2014}. The authors} \edit2{analyzed} \edit1{the CME with the two-stage acceleration (both works studied the same event). In their works, each stage of the acceleration was accompanied by the X-ray flux peak. They interpreted the second X-ray peak as a second stage of magnetic reconnection, which caused the second acceleration phase.}

\edit1{\citet{Gosain2016} reported another CME with a two-stage acceleration. The authors calculated the decay index of the coronal magnetic field and found that there were two altitude ranges} \edit2{for} \edit1{the prominence stability. They concluded that the two areas of stability caused} \edit2{a} \edit1{ two-stage acceleration of the CME.}

\edit2{The} \edit1{above-mentioned papers reported two-stage acceleration low in the corona. In this work, the second impulsive acceleration phase occurred at a height of several solar radii. This is a notable observation that deserves discussion.}

\edit3{During the first impulsive acceleration phase, the two-ribbon flare occurred. The speed of the ribbon separation and the derivative of the X-ray flux (signatures of the flare reconnection) peaked at the beginning of the first impulsive acceleration phase (see Figure~\ref{F:mg_flux}). We think that the first acceleration phase was caused by the flare reconnection described by the standard CME model \citep[CSHKP model, ][]{Carmichael1964, Sturrock1966, Hirayama1974, Kopp1976}.}

\edit3{We believe that the temporal correlation between the first acceleration phase and signatures of the flare reconnection is not a coincidence. For the majority of CMEs, the impulsive acceleration is correlated with the rise of the soft X-ray flux \citep{Marivcic2007, Bein2012}, the hard X-ray bursts \citep{Temmer2008, Temmer2010}, and peak of the reconnection rate \citep{Qiu2004}.}

\edit3{There are no cause-effect relationship between flares and CMEs. Both phenomena result from the same complex magnetic process \citep{Webb2012, Chen2011}. The flare reconnection increase the CME acceleration, and faster motion of the CME increase the reconnection rate \citep{Schmieder2015}.}

\edit1{The second impulsive acceleration phase was preceded by the deceleration phase. Right before the deceleration phase, the studied CME collided with another CME. We think that both the deceleration and acceleration  were caused by the complex dynamics of the CME-CME interaction: deceleration corresponds to CME compression,} \edit2{while} \edit1{acceleration corresponds to CME restitution. \citet{Shen2012a} performed MHD simulation of the CME-CME collision and showed that, during the collision, the CME experiences both acceleration and deceleration.}

\edit1{It may look strange that the studied CME increased its velocity after the collision. However, the CME-CME collision is a complex interaction of two magnetic plasma clouds that occurs in 3D. Furthermore, CME-CME collision could be super-elastic \citep{Shen2012b, Shen2013}. Therefore, it is possible that the studied CME could increase its speed as a result of the collision.}

\edit2{The} \edit1{CME-CME collisions observed in the heliosphere} \edit2{lasted} \edit1{for 8--24 hours. This duration roughly coincides with the Alfven crossing time of a CME \citep{Lugaz2017}.} 

\edit2{We estimate the CME-CME collision time ($\tau$) as a total duration of the impulsive deceleration and the second impulsive acceleration phases ($\tau \approx$~100~min).} 
\edit1{The corresponding Alfven speed ($v_A$) should be equal:}
\begin{equation}
v_A \approx \frac{L}{\tau} \approx 250 \ \text{km s$^{-1}$,}
\end{equation}
\edit2{where $L$ is the size of the studied CME during the collision ($L \approx$~2~$R_\odot \approx$~1500~Mm). The value of $v_a$} \edit1{coincides within the order of magnitude with the Alfven speed inside CMEs.}

\edit1{During the CME collision, the acceleration profiles of the CME core and frontal loop look similar: they both initially impulsively decelerate and then impulsively accelerate. However, the values of the acceleration differ. This means that different parts of the CME react differently to the collision.}

\subsection{CME Core and a Prominence}

In white-light coronagraph images, approximately 30\% of  CMEs have a three-part structure: \edit2{a} bright core, \edit2{a} dark cavity, and \edit2{a} bright frontal loop \citep{Illing1985, Webb1987}. It is possible that more CMEs have a three-part structure, but we do not see it \edit2{due to projection effects}.

\citet{Illing1986} compared white-light and $H_\alpha$ images of the \textit{SMM} coronagraphs \citep{Hundhausen1999}, and showed that several CMEs contained cold prominence plasma inside their cores. In white-light coronagraphs images, the CME core often looks like a prominence \citep[\edit2{a} long linear twisted structure;][]{Plunkett2000}.  Since the core resembles the prominence and some CME cores contained \edit2{cool} prominence plasma, the core is usually interpreted as \edit2{a cool} erupting prominence \citep{House1981, Webb2012, Parenti2014}.

However, a CME core could have a wide range of temperatures: cool \citep[0.03--0.3~MK; ][]{Akmal2001, Ciaravella1997, Ciaravella1999, Ciaravella2000}, warm \citep[$\approx$1~MK; ][]{Ciaravella2003, Landi2010, Reva2016b}, or hot  \citep[5--10~MK; ][]{Reeves2011, Song2014, Nindos2015}. The warm and hot plasma inside the CME cores could be interpreted in two ways: the prominence is heated during the eruption \citep{Filippov2002} or the CME core is not a prominence. As we see, the relationship between \edit2{the} core and the prominence is unclear.

In this work, we studied a CME with a three-part structure in the LASCO images. Its core corresponded to the fluxrope, \edit2{as} observed in the Fe~171~\AA\ images. However, TESIS He~304~\AA\ images showed that the prominence drained down, before the CME reached the LASCO field of view. 

\edit2{The apparent drainage of the prominence could be explained in several ways. Firstly, the prominence could actually drain down. In this case, the core would not contain the prominence plasma. Probably, the core observed in LASCO images corresponded to the warm ($\approx$1~MK) fluxrope that surrounded the prominence.} 

\edit2{Secondly, it is possible that the small part of the prominence escaped with the CME, but we did not observe this, due to the limited cadence or sensitivity of the instruments. In this case, the core would correspond to the escaped part of the prominence.}

\edit2{Finally, the prominence could be heated during the eruption in such a way that it creates the illusion of drainage. In this case, the core would contain not a cool prominence plasma, but rather a heated prominence plasma.}

\edit1{Although the CME core in the white-light images is often interpreted as a cold erupting prominence, our observations} \edit2{suggest} \edit1{that this assumption is not always correct.} \edit2{Evolution of the erupting prominence in the distance range of 1.2--2~$R_\odot$ requires further investigation.}

\section{Conclusion}

\edit1{We studied the CME that had a complex acceleration profile: two impulsive acceleration phases and one impulsive deceleration phase. The CME showed signatures of several acceleration mechanisms: kink instability, mass drainage, flare reconnection, and CME-CME collision.}

\edit1{The kink instability triggered the CME eruption. The prominence drainage happened after the kink instability. The drainage played an auxiliary role: it decreased the CME mass, which helped its acceleration.}

\edit2{The first impulsive acceleration phase was caused by the flare reconnection below the CME.} During the first phase, \edit2{the two-ribbon flare occurred and the SphinX X-ray flux increased.}

\edit1{The second impulsive acceleration and a deceleration phase were caused by the CME-CME interaction. We think that the deceleration corresponds to CME compression, and the acceleration to CME restitution.}

The studied event shows that CMEs are complex phenomena that cannot be explained \edit2{by} only one acceleration mechanism. We should seek a combination of different mechanisms that accelerate CMEs at different stages of their evolution.

\acknowledgments
This research was supported by the Russian Science Foundation (project No. 17-12-01567).

\bibliographystyle{aasjournal}
\bibliography{mybibl}

\end{document}